\documentclass[preprint,aps,showpacs,showkeys,tightenlines,superscriptaddress]{revtex4}
\usepackage{epsfig}
\newcommand{\vi}[1]{\mbox{\boldmath $#1$}}
\newcommand{\vis}[1]{\mbox{\boldmath ${\scriptstyle #1}$}}
\begin{document}
\title{$^{22}$C: An $S$-Wave Two-Neutron Halo Nucleus}
\author{W. Horiuchi}%
\affiliation{Graduate School of Science and Technology, 
Niigata University, Niigata 950-2181, Japan}
\author{Y. Suzuki}
\affiliation{Department of Physics, and Graduate 
School of Science and Technology, Niigata University, Niigata
950-2181, Japan}

\begin{abstract}
A dripline nucleus $^{22}$C is studied in a Borromean three-body model of 
$^{20}$C+$n$+$n$. The valence neutrons, interacting via a realistic 
potential, are constrained to be orthogonal to the occupied 
orbits in $^{20}$C. 
We obtain ample results supporting that $^{22}$C is an ideal $S$-wave 
two-neutron halo nucleus: The ground state is bound by 390-570 keV, 
the  root mean square neutron and proton radii are 4.0 
and 2.4 fm, and the two neutrons are 
predominantly in $(s_{1/2})^2$ orbits. The binding mechanism of $^{22}$C 
is discussed. One- and two-body 
density distributions elucidate the halo character as well as 
the correlated motion of the neutrons. 
The reaction cross sections of $^{22}$C+$^{12}$C collisions are predicted. 
\end{abstract}
\maketitle
\draft

The subshell closure of $N$=$14$ and $N$=$16$ is one of the topics 
discussed intensively in the study of neutron-rich 
nuclei, and the $N$=$14$ closure has experimentally been confirmed 
around $^{22}$O~\cite{ozawa00,thirolf,cortina,stanoiu,becheva}. This issue  
is closely related to the competition of $0d_{5/2}$ 
and $1s_{1/2}$ neutron orbits. In fact, they play  a 
vital role in determining the ground state structure of 
$A$=$15$--20 carbon isotopes. For example, the ground state of 
$^{16}$C is found to contain 
the $(1s_{1/2})^2$ and $(0d_{5/2})^2$ 
configurations nearly equally~\cite{yamaguchi,horiuchi}, whereas the 
last neutron in $^{19}$C is in the $1s_{1/2}$ orbit, 
forming one-neutron halo structure~\cite{nakamura,maddalena}. 
The deformations of carbon isotopes are discussed to have a  
strong $N$-dependence~\cite{kanada,sagawa}. An overview of the 
structure of carbon isotopes is given in Ref.~\cite{brown}.

No information is available to determine whether the $N$=$14$ 
subshell closure occurs in $^{20}$C. The systematics of the interaction 
cross section suggests, however, that the radius of $^{20}$C is 
smaller than that of 
$^{19}$C~\cite{ozawa}, so it is natural to assume   
that the ground state of $^{20}$C predominantly consists of 
a $(0d_{5/2})^6$ configuration. If its dominant component were 
$(0d_{5/2})^4(1s_{1/2})^2$, 
one more neutron could be added to the $0d_{5/2}$ orbit to 
form a particle-stable $^{21}$C, which is in contradiction to observation.  

In this study we will demonstrate that $^{22}$C is an 
$s$-wave two-neutron halo nucleus on the basis of the analysis of its 
structure including the neutron and proton densities.  For $Z \le 8$,
$^{22}$C is an only dripline nucleus which the interaction or reaction 
cross section 
measurement on a $^{12}$C target has not reached yet~\cite{ozawa}, so it is of 
interest for a future measurement to predict the reaction cross 
section of $^{22}$C. The neutron and proton densities obtained here will
be useful to estimate the reaction cross section of $^{22}$C on a
proton target, which is being investigated experimentally~\cite{tanaka}.  
Our model is that $^{22}$C is a three-body system of $^{20}$C+$n$+$n$, 
and that $^{20}$C has the $(0d_{5/2})^6$ configuration.  
$^{22}$C is thus Borromean, just as $^{11}$Li is. Though 
$^{22}$C may be expected to be much like $^{11}$Li in its halo character, 
a remarkable difference will show up: 
In $^{11}$Li both of $(0p_{1/2})^2$ and $(1s_{1/2})^2$ components 
contribute to producing its halo~\cite{thompson,simon,varga}, 
whereas in $^{22}$C 
only an $(s_{1/2})^2$ 
component will be predominant. Another difference to be noted is that 
the $^{20}$C core has zero spin, which will make the content of angular 
momentum coupling in $^{22}$C simpler than that in $^{11}$Li.

The wave function for $^{22}$C is determined from the 
following Hamiltonian 
\begin{equation}
H=T_{\vis \lambda}+T_{\vis \rho}+U_1+U_2+v_{12},
\label{hamiltonian}
\end{equation}
where the subscripts, ${\vi \lambda}$ and ${\vi \rho}$, of the 
kinetic energies stand for 
the relative distance vectors of the three-body system. 
The two-neutron potential $v_{12}$ 
is taken from the realistic G3RS (case 1) potential~\cite{tamagaki} which 
contains central, tensor and spin-orbit forces 
and reproduces the nucleon-nucleon scattering data 
as well as the deuteron properties. $U_i$ is the 
$n-^{20}$C potential whose form is assumed as  
\begin{equation}
U=-V_0f(r)+V_1{\vi \ell}\cdot{\vi s}{\frac{1}{r}}{\frac{d}{dr}}f(r)+
V_s{\rm e}^{-\mu r^2}{\cal P}_s,
\label{opt.pot}
\end{equation}
where $f(r)$=$[1+{\rm exp}({\frac{r-R_{\rm c}} {a}})]^{-1}$ with 
$R_{\rm c}$=$r_0A_{\rm c}^{\,1/3}\, (A_{\rm c}\!=\!20)$. 
The operator ${\cal P}_s$ of the last term projects to the $s$ wave of  
the $n-^{20}$C relative motion, so this term modifies the $s$-wave 
potential strength. To determine the parameters of $U$, we 
take into account the conditions that (i) the $1s_{1/2}$ orbit is 
unbound as $^{21}$C is unstable for a neutron emission, 
and (ii) the $0d_{5/2}$ orbit is bound by at most 2.93 MeV, which is   
the neutron separation energy of $^{20}$C. 
Without the ${\cal P}_s$ term, the above conditions were barely met only by 
making $V_1$ much larger than the standard strength~\cite{bm}. 
The set A potential in Table~\ref{c20-n.pot} corresponds to this case.  
With the ${\cal P}_s$ term included, 
we have more freedom to generate different potentials, which offer the 
opportunity of investigating the sensitivity of $U$ on theoretical results. 
The spin-orbit strength $V_1$ is fixed to be the standard value.  
Three sets of $U$ of this type are listed in 
Table~\ref{c20-n.pot} as B, C and D. These potentials are 
determined by giving 
different values for the $0d_{5/2}$ single-particle (s.p.) 
energy: Set B potential gives the deepest energy, while set D the shallowest 
energy. The energies of the lower s.p. orbits turn out to be considerably 
different. It should be noted, however, that our result for $^{22}$C 
never depends on these energies but on their s.p. wave
functions as will be seen later. Fortunately, the different potentials 
chosen here give almost 
the same s.p. wave function for each occupied orbit. All of the potentials 
are set to predict the $1s_{1/2}$ s.p. energy almost zero. It may 
be probable that the $s$-wave potential 
strength is further weaker. In that case, 
the ground state energy of $^{22}$C which we will obtain below is to be 
considered a minimum.

\begin{table}[t]
\caption{Parameters of the $n-^{20}$C potential $U$. $\mu\!=\!0.09$ fm$^{-2}$. 
$a$ and $r_0$ are 0.6 fm and 1.3 fm for set A, while they are 0.65 fm and 
1.25 fm for sets B, C and D. $\varepsilon$ is the s.p. energy 
of the $n-^{20}$C relative motion. Energy and length are given in units of MeV and 
fm, respectively.}
\label{c20-n.pot}
\begin{center}
\begin{tabular}{cccccccccc}
\hline\hline
&  &$V_0$ & $V_1$ & & $V_s$ & $\varepsilon(0s_{1/2})$ 
& $\varepsilon(0p_{3/2})$  & $\varepsilon(0p_{1/2})$ & $\varepsilon(0d_{5/2})$  \\
\hline
set A & &33.22   & 42.10 & & 0.00 & $-$19.03 & $-$9.86  & $-$4.77 & $-$1.00 \\
set B & &43.24   & 25.63 & & 9.46 & $-$19.79 & $-$14.32  & $-$11.00 & $-$2.93 \\
set C & &41.08   & 25.63 & & 7.14 & $-$19.56 & $-$12.88  & $-$9.58 & $-$1.93 \\
set D & &38.76   & 25.63 & & 4.66 & $-$19.31 & $-$11.37  & $-$8.09 & $-$0.93 \\
\hline\hline
\end{tabular}
\end{center}
\end{table}

The ground state of $^{22}$C is described as follows: 
\begin{equation}
\Psi=\Phi_{\rm c}\Phi_{2n},\ \  \ {\rm with}\ \ 
\Phi_{2n}=\sum_{i=1}^KC_i \Phi(\Lambda_i,A_i),
\label{totalwf}
\end{equation}
where $\Phi_{\rm c}$ is the intrinsic wave function of 
$^{20}$C and the valence neutron part $\Phi_{2n}$ is given as 
a combination of correlated Gaussian bases 
\begin{equation}
\Phi(\Lambda,A)\!=\!(1-P_{12})  \left\{ {\rm e}^{-{\frac{1}{2}}
\tilde{\vis x}A{\vis x}} [[{\cal Y}_{\ell}({\vi x}_1)
{\cal Y}_{\ell}({\vi x}_2)]_L\chi_S(1,2)]_{00}\right\},
\label{base}
\end{equation}
where $P_{12}$ permutes the neutron coordinates and 
$\tilde{\vi x}A{\vi x}= 
A_{11}{\vi x}_1^{\, 2}$+$2A_{12}{\vi x}_1
\!\cdot\!{\vi x}_2$+$A_{22}{\vi x}_2^{\, 2}$. The coordinates 
${\vi x}_1$=${\vi \rho}$+${\frac{1}{2}}{\vi \lambda}$ and 
${\vi x}_2$=${\vi \rho}\!-\!{\frac{1}{2}}{\vi \lambda}$ 
are the distance vectors of the neutrons from the 
center of mass (c.m.) of $^{20}$C. The angular parts of the two-neutron 
motion are described using 
${\cal Y}_{\ell m}({\vi r})\!=\!r^{\ell}Y_{\ell m}(\hat{\vi r})$ and 
they are coupled with the spin part $\chi_S$ to the total angular 
momentum zero. The basis function is specified by a set of angular 
momenta $\Lambda$=$(\ell,S)$ ($L$=$S$), and a 2$\times$2 
symmetric matrix $A$ ($A_{21}$=$A_{12}$). 
The two neutrons are explicitly correlated due to the term 
$A_{12}{\vi x}_1$$\cdot$${\vi x}_2$, the inclusion of which assures a precise 
solution in a relatively small dimension~\cite{svm}. 

It is vital to take into account the Pauli principle for 
the valence neutrons in determining  
the energy and corresponding wave function. 
Though the fulfillment of antisymmetrizing the core and valence 
neutrons is beyond the present model,  
the Pauli constraint is included 
by imposing that the valence neutrons cannot occupy 
any s.p. orbits $u_{n\ell jm}$ of $\Phi_{\rm c}$. Here  
$u_{n\ell jm}$ are 
generated from $U$, and $n\ell j$ runs over 
$0s_{{1}/{2}},\, 0p_{{3}/{2}}$, $0p_{{1}/{2}}$, and $0d_{5/2}$. 
We used the stochastic 
variational method (SVM)~\cite{svm} to optimize the 
parameter matrices $A$. The SVM increases the basis dimension 
one by one by testing a number of candidates which are chosen 
randomly. The basis selection with the SVM is very 
effective for taking care of the short-range 
repulsion of $v_{12}$ as well as satisfying the orthogonality constraint. 

The most important channel for the binding of $^{22}$C was found to be 
$\Lambda$=$(0,0)$, and other channels 
included were $(1,0)$, $(2,0)$, $(1,1)$, and 
$(2,1)$. Note, however, that our correlated 
basis functions in practice include higher partial waves as well. 
Convergent results are obtained with the basis 
dimension of $K\!\approx \!300$. The $U$-dependence of the solution 
is moderate as shown in Table~\ref{c22}. The result with the 
$\ell$-independent 
set A potential is similar to those with the 
other potentials, especially set D potential. 
This indicates that the present result is not very sensitive to 
the potential provided that it is chosen to satisfy the two conditions. 
The ground state energy is about $-$390 to $-$570 keV with respect to 
the $^{20}$C+$n$+$n$ threshold, which is consistent with 
the empirical value of $-0.423 \pm 1.140$ MeV~\cite{audi}. 
In order to
see the importance of both spatial and angular correlations of the 
basis functions, we repeated the following calculations. The first was to 
include only the single channel of $\Lambda$=$(0,0)$, and then 
the ground state energy turned out to be $-$0.29 MeV for set B. 
In the second calculation 
which truncates the basis functions to those with $\Lambda$=$(0,0)$ and 
$A_{12}$=$0$ (no correlation calculation), we obtained the result that 
the ground state is 
bound by at most 90 keV. Thus the inclusion of the correlated bases is 
found to gain the energy of about 400 keV.

\begin{table}[t]
\caption{Properties of $^{22}$C. Length is given in units of fm.}
\label{c22}
\begin{center}
\begin{tabular}{ccccccccccccc}
\hline\hline
&  &$E$(MeV) & $R_{\rm rms}^{n}$ &$R_{\rm rms}^{p}$ &$R_{\rm rms}^{m}$ & & $\sqrt{\langle{\vi x}_1^2\rangle}$ & $\sqrt{\langle {\vi \rho}^2\rangle}$  
&$\sqrt{\langle {\vi \lambda}^2\rangle}$ & $\langle {\vi x}_1$$\cdot$ $ {\vi x}_2\rangle$ 
& $P_{S=0}$ & $\langle(s_{1/2})^2\rangle$ \\
\hline
set A & &$-$0.413&4.11 &2.44 & 3.73 & & 8.17 & 6.19 & 10.7 & 9.80 & 0.998 & 0.968 \\
set B & &$-$0.489&3.96 &2.43 & 3.61 & & 7.54 & 5.86 & 9.48 & 11.9 & 0.981 & 0.915 \\
set C & &$-$0.573&3.93 &2.43 & 3.58 & & 7.37 & 5.66 & 9.44 & 9.83  & 0.990 & 0.942 \\
set D & &$-$0.388&4.12 &2.44 & 3.74 & & 8.21 & 6.29 & 10.6 & 11.6 & 0.995 & 0.954 \\
\hline\hline
\end{tabular}
\end{center}
\end{table}

The rms neutron, proton and matter radii of $^{22}$C, assuming pointlike 
nucleons, are listed in Table~\ref{c22}. They are 
obtained using the corresponding radii of $^{20}$C, 3.23, 2.37 and 2.99 
fm, which are calculated from $\Phi_{\rm c}$.  
The rms neutron radius is  
3.9--4.1 fm. The rms matter radius results in about 3.6--3.7 fm, which 
corresponds to that 
of a stable nucleus with $A\!\approx \!60$. Accordingly one may call 
$^{22}$C a giant halo nucleus. 
The probability of finding the spin-singlet neutrons, $P_{S=0}$, shows  
that the ground state of $^{22}$C almost consists of the $S$=$0$ 
component. Therefore, the non-central potentials have small  
expectation values and play a minor role in binding $^{22}$C:  
In set B case, the value of $\langle v_{12}({\rm tensor+spin}$-${\rm orbit})\rangle $ 
is only 7 keV and that of $\langle U_{1}({\rm spin}$-${\rm orbit})\rangle $ 
is 57 keV. Thus the binding energy contribution virtually comes from 
the kinetic energy and the central potentials of both $U$ and 
$v_{12}$. 

It is interesting to understand how the Borromean system is bound. 
First of all, we note that the non-central forces are found to give 
negligible contributions. Rewriting the kinetic energy as 
$T_{\vis \lambda}+T_{\vis \rho}=T_1+T_2+T_{\rm rc}$~\cite{suzuki88}, 
where $T_i$ is the kinetic energy for the $n-^{20}$C relative motion and 
$T_{\rm rc}$ is the recoil correction term, we decompose the energy contribution 
as follows:
\begin{equation}
E=2\langle T_1+U_1 \rangle+ \langle T_{\rm rc} \rangle +\langle v_{12} \rangle.
\end{equation}
The decomposition for set B 
case is $2\!\times\!(7.185-6.436)-0.118-1.868=-0.489$ MeV. Except for 
the small contribution of the 
$\langle T_{\rm rc} \rangle $ term, we conclude that  
the binding of $^{22}$C is obtained by  
a delicate balance of the two factors: 
One is that the attraction of $v_{12}$, though not large, 
keeps the neutrons from separating, and 
the other is the weak attraction of $U$ which puts the neutron in continuum.  

We calculate the probability $\langle(\ell_j)^2\rangle$ of finding the 
halo neutrons in an $(\ell_j)^2$ component. 
The $\langle(s_{1/2})^2\rangle$ value is listed  
in Table~\ref{c22}. Other probabilities are, for set B case,  
0.033, 0.024, 0.009, 0.007, 0.003, 0.003 for 
$\ell_j$=$d_{3/2},\, p_{3/2},\, p_{1/2},\, f_{7/2},\, d_{5/2},
\, f_{5/2}$, 
respectively. The other potential sets give similar results. 
We find that the $(s_{1/2})^2$ component is predominant and 
many other components have small admixtures.  
Since no bound orbit exists for the valence neutron, 
a realistic shell-model description taking into account these 
components would be hard. 
On the contrary, the present approach has the advantage that it 
requires no s.p. energies, no matter how high the valence 
neutrons are excited. 

\begin{figure}[b]
\epsfig{file=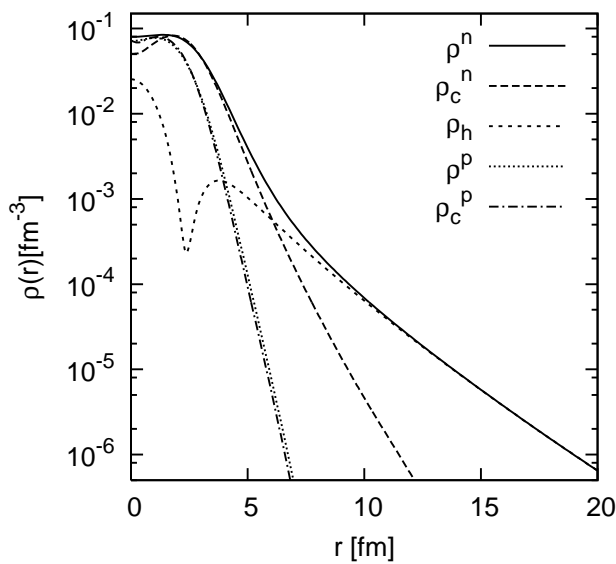,width=11.0cm,height=7.7cm}
\caption{The neutron and proton densities of $^{22}$C and
 $^{20}$C. $\rho_{\rm h}$ is the halo-neutron density. Set B potential 
is used.} 
\label{cdens}
\end{figure}

The halo behavior of $^{22}$C is exhibited through  
the neutron density, $\rho^{n}({\vi r})$, which is given by 
\begin{equation}
\rho^{n}({\vi r})=\langle \Phi_{2n} \mid \rho_{\rm c}^{n}(
\textstyle{\frac{2}{22}}{\vi \rho}
+{\vi r})\mid \Phi_{2n} \rangle+ \rho_{\rm h}({\vi r}), 
\label{density}
\end{equation}
where $\rho_{\rm c}^{n}({\vi r})$ stands for the intrinsic neutron density 
of $^{20}$C, which is calculated from $\Phi_{\rm c}$, and $\rho_{\rm h}({\vi r})$ 
is the halo-neutron density with respect to the c.m. of $^{22}$C 
\begin{equation}
\rho_{\rm h}({\vi r})=\langle \Phi_{2n} \mid \sum_{i=1}^2
\delta({\vi x}_i-\textstyle{\frac{2}{22}}{\vi \rho}-{\vi r})
\mid \Phi_{2n} \rangle.
\end{equation}
The integration of $\rho_{\rm c}^{n}$ in Eq.~(\ref{density}) takes care of 
the fluctuation of the c.m. of $^{20}$C around the c.m. of $^{22}$C.  
The proton density is given by 
\begin{equation}
\rho^{p}({\vi r})=\langle \Phi_{2n} \mid 
\rho_{\rm c}^{p}(\textstyle{\frac{2}{22}}{\vi \rho}
+{\vi r})\mid \Phi_{2n} \rangle. 
\end{equation}
These densities are displayed in 
Fig.~\ref{cdens}. The contribution of the halo density to $\rho^{n}$ exceeds 
that of the core density beyond $r$=$6.2$ fm. Note that 
$\rho_{\rm h}({\vi r})/2$ is, roughly speaking, the squared single 
halo-neutron wave function. The dip at 
around $r$=$2.4$ fm is due to the orthogonality 
of $\Phi_{2n}$ to the $0s_{1/2}$ orbit. 

\begin{figure}[t]
\epsfig{file=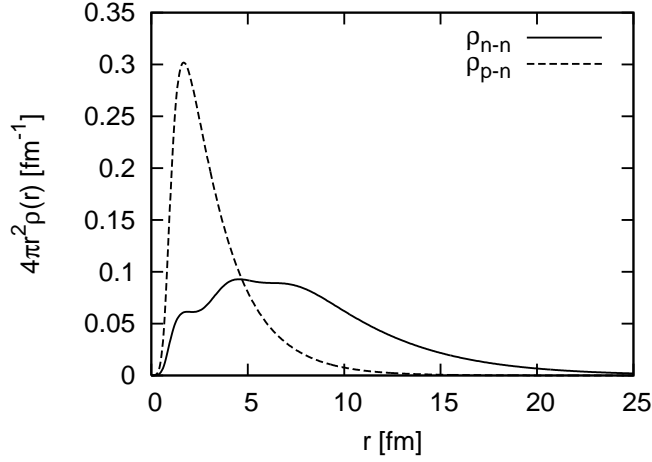,width=9.0cm,height=6.3cm}
\caption{Comparison of the two-body density distribution of 
the halo neutrons in $^{22}$C with that of the proton and neutron in 
the deuteron. Set B potential is used.}
\label{nnrel.dens}
\end{figure}

It is of interest to examine the correlated motion of the two neutrons.   
A two-neutron subsystem with $S=0$ is often called a di-neutron 
when they have spatial extension comparable to that of the deuteron. 
The two-body halo-neutron distribution function,  
\begin{equation}
\rho_{n-n}({\vi r})=\langle \Phi_{2n}\mid \delta({\vi \lambda}-{\vi r})\mid 
\Phi_{2n}\rangle, 
\end{equation}
is compared in Fig.~\ref{nnrel.dens} with the corresponding $p-n$ 
distribution function of the deuteron 
\begin{equation} 
\rho_{p-n}({\vi r})=\frac{1}{3}\sum_{M=-1}^{1} \langle 
\Phi_{d}({1M})\mid \delta({\vi r}_{p}-{\vi r}_{n}-{\vi r})\mid 
\Phi_{d}({1M})\rangle,
\end{equation} 
which is calculated using the G3RS potential. 
It is found that $\rho_{n-n}({\vi r})$ has a distribution much wider 
than $\rho_{p-n}({\vi r})$. Thus the di-neutron correlation is not 
prominent in $^{22}$C. The value of 
$\langle T_{\vis \lambda}\!+\!v_{12}\rangle $ is 
6.16 MeV (set B), which is to be compared to $-$2.28 MeV (G3RS) of 
the deuteron. 
Since $\rho_{n-n}({\vi r})$ has a long tail, one may expect 
that the use of a two-nucleon potential with a one-pion exchange 
tail would give a potential 
energy different from the G3RS potential of a Gaussian tail. 
To check this point, we estimated the energy 
difference arising when the singlet-even central potential of  
G3RS is replaced with that of the OPEG potential 
(case 1)~\cite{tamagaki}, using 
\begin{equation}
\int d{\vi r} \rho_{n-n}({\vi r}) \Big[v_{12}({\rm OPEG})-v_{12}({\rm
 G3RS})\Big]
\end{equation}
and found that the energy gain is only 6 keV. 

\begin{figure}[t]
\epsfig{file=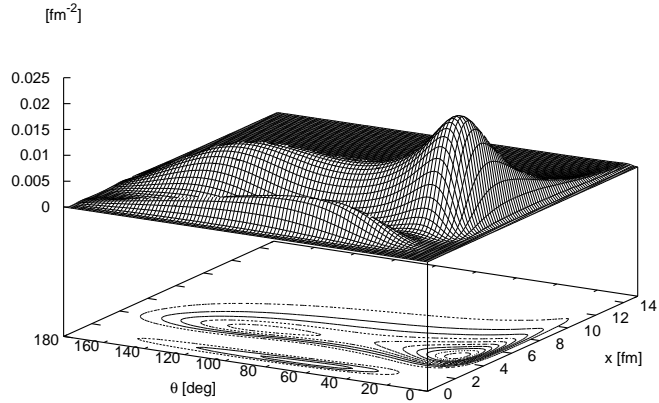,width=9.0cm,height=5.85cm}
\caption{The two-neutron correlation function $\rho(x,x,\theta)$ 
weighted by $8\pi^2x^4{\rm sin}\theta$. The lower panel is its contour map. 
Set B potential is used.}
\label{t.p.dens}
\end{figure}

Another function of interest is the two-neutron correlation 
function defined as  
\begin{equation}
\rho(x_1,x_2,\theta)=\langle \Phi_{2n} \mid
 \Phi_{2n} \rangle_{\rm spin},
\end{equation}
where 
$\theta$ is the angle between ${\vi x}_1$ and ${\vi x}_2$ 
and $\langle\cdots \rangle_{\rm spin}$ indicates that the 
integration is to be done over the spin coordinates only. 
Figure~\ref{t.p.dens} displays the value of 
$8\pi^2x^4{\rm sin}\theta\,\rho(x,x,\theta)$.  
One prominent peak appears around $x$=$5.0$ fm and $\theta$=$17^{\circ}$, 
which is often attributed to the correlation of di-neutron type, but 
the spatial extension of the two neutrons is too wide to be called the 
di-neutron, as shown in Fig.~\ref{nnrel.dens}. 
The peak is followed by a plateau extending to larger angles. The valley  
of the correlation function which appears at around $x$=$2.4$ fm 
reflects the dip observed in the halo-neutron density of Fig.~\ref{cdens}.

\begin{figure}[t]
\epsfig{file=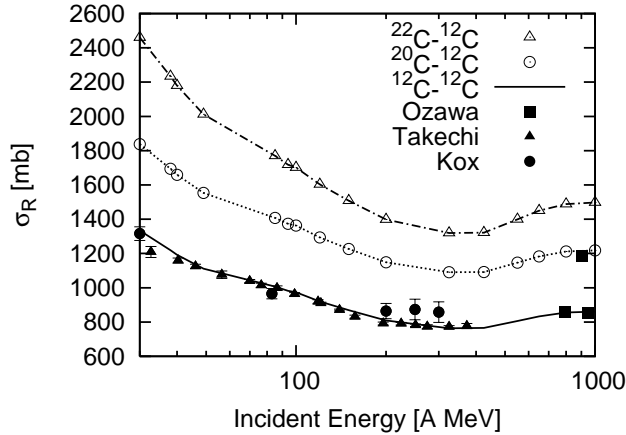,width=8.5cm,height=5.95cm}
\caption{Prediction of the reaction cross sections of $^{22}$C and
 $^{20}$C on a $^{12}$C target. Experimental data are taken from 
Refs.~\cite{ozawa,fukuda,kox}. }
\label{ccreac}
\end{figure}

The interaction cross section data for 
the carbon isotopes are available up to $^{20}$C for high incident 
energies~\cite{ozawa}. 
With a reaction model proposed in Ref.~\cite{utility}, we 
predict the reaction cross section $\sigma_{\rm R}$ 
of $^{22}$C (and $^{20}$C) using the calculated densities. 
To make the prediction reliable, we modify a nucleon-nucleon ($NN$) 
profile function $\Gamma_{NN}$ available in literatures~\cite{ray} 
so as to 
reproduce both the elastic scattering cross section and the total cross 
section of the $NN$ collision. Details will be published elsewhere. 
Figure~\ref{ccreac} displays  
$\sigma_{\rm R}(^{12,20,22}{\rm C})$ 
on a $^{12}$C target calculated at several incident energies. 
A good agreement between theory and experiment for 
$\sigma_{\rm R}(^{12}{\rm C})$ confirms the validity of 
the modification of $\Gamma_{NN}$. 
The $\sigma_{\rm R}(^{20}{\rm C})$ value at the incident 
energy of 900 $A \, {\rm MeV}$  
is fairly well reproduced, which indicates that our model for 
$\Phi_{\rm c}$ is acceptable at least in its prediction for 
the radius of $^{20}$C. We thus expect that 
$\sigma_{\rm R}(^{22}{\rm C})$, or at 
least the increase of the cross sections, $\sigma_{\rm R}(^{22}{\rm C})
-\sigma_{\rm R}(^{20}{\rm C})$, is predicted to good approximation. 
A measurement of $\sigma_{\rm R}(^{22}{\rm C})$ for a wide range 
of incident energies will provide us with valuable information for 
quantifying the extent to which the halo reaches in far distances. 
We are studying the reaction cross section of $^{22}$C on a proton
target as well. 

To conclude, we studied the ground state structure of $^{22}$C in the 
$^{20}$C+$n$+$n$ three-body model with the orthogonality constraint. The 
$N$=$14$ subshell closure was assumed for $^{20}$C. We showed  
that $^{22}$C is an almost pure $S$-wave two-neutron 
halo nucleus, and that the non-central forces play no active role 
in binding this fragile system. A measurement of the reaction cross 
section of $^{22}$C+$^{12}$C is desired to establish the halo 
structure experimentally.

\vspace{3mm}

We thank A. Kohama for his interest and valuable discussions
and M. Fukuda and M. Takechi for 
sending us the reaction cross section data of $^{12}$C+$^{12}$C. 
This work was in part supported by 
a Grant for Promotion of Niigata University 
Research Projects (2005-2007).

\end{document}